\input stromlo

\title A Nuclear Disk and massive BH in NGC 4342.

\shorttitle NGC 4342; An E/S0 with Nuclear Disk and Black Hole.

\author Frank C. van den Bosch and Walter Jaffe

\shortauthor van den Bosch \& Jaffe

\affil Leiden Observatory, The Netherlands


\def\etal{et al.~}
\def\Msun{{\rm\,M_\odot}}
\def\kms{{\rm\,km\,s^{-1}}}
\def\pc{{\rm\,pc}}

\def\Mpc{{\rm\,Mpc}}


\index{NGC 4342}
\index{nuclear disks}
\index{massive BHs}

\abstract We discuss photometric and spectroscopic data of the E/S0 galaxy 
NGC 4342, obtained with the {\it Hubble Space Telescope} (HST). This galaxy 
harbors, in addition to its outer disk, a very small, stellar disk in its 
nucleus. Jeans-modeling suggests the presence of a 3-6$\times 10^8 \Msun$ 
black hole (BH) in the nucleus. This galaxy therefore deviates significantly 
from the correlation between BH mass and bulge mass, found for other BH 
candidate galaxies.

\section Introduction

The discovery with the HST of
a number of early-type galaxies with very small, stellar disks
with scalelengths of $\sim 20 \pc$
(e.g. van den Bosch \etal 1994; Lauer \etal 1995) has opened 
new windows on galaxy dynamics and formation. From a dynamical point
of view nuclear, stellar disks are interesting since their kinematics
allows an accurate determination of the central mass density of these
galaxies (van den Bosch \& de Zeeuw 1996). 
An interesting question concerns the origin of these nuclear disks.
Did they form coeval with the host galaxy, or did they arise from secular 
evolution? In the latter case, gas infall to the center, induced by either 
a bar or a merger, could, after subsequent starformation, result in the
nuclear disk.

In order to study the dynamics of the nuclear disks, and to address the
question of their formation, we have obtained both multi-color photometry 
and spectroscopy with the HST of two E/S0s in the Virgo cluster: NGC 4342 and
NGC 4570. Here we focus on NGC 4342, and discuss some simple 
dynamical modeling. The full data on both galaxies will be published in
a forthcoming paper (van den Bosch \& Jaffe 1997). 

\section Multi-color photometry with the HST 

We have obtained $U$, $V$, and $I$-band images of NGC 4342 
with the {\tt WFPC2} aboard the HST. Contour maps at two different scales
of the $V$-band image are shown in Fig. 1. They clearly reveal the 
multi-component structure. The panel on the right reveals the presence of
the nuclear disk by the lemon-shape of its isophotes.
Proper dynamical modeling requires an adequate (luminosity)-mass model 
of the galaxy, that after projection accurately fits the observed surface 
brightness. We have used the Multi Gaussian Expansion (MGE) method developed 
by Monnet, Bacon \& Emsellem (1992). This method is particularly well suited 
to fit complex surface brightness distributions.
In Fig. 1 we have overplotted the projected surface brightness of our
MGE model that after convolution with the HST PSF best fits the V-band image.

\figureps[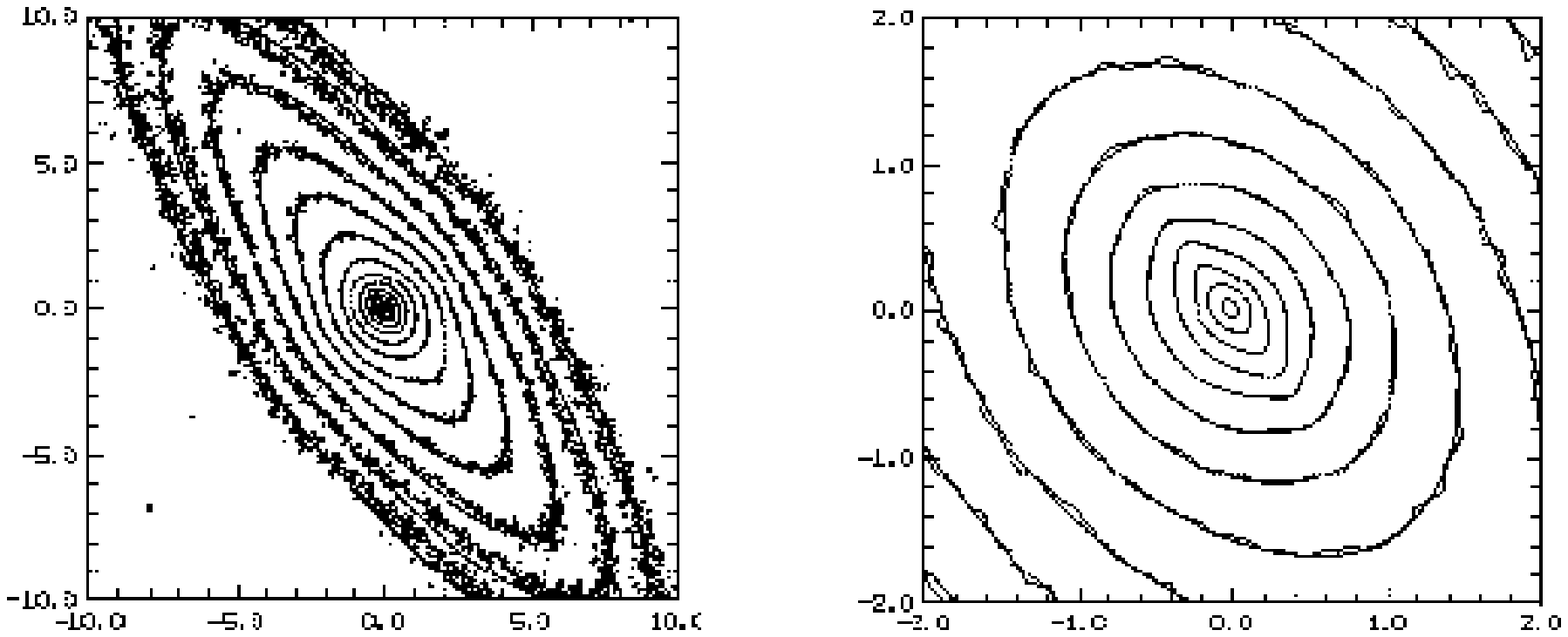,.9\hsize] 1. Contour maps of the WFPC2 V-band images
of NGC 4342 and NGC 4570 at two different scales: $20"\times 20"$ and
$4"\times4"$. Overplotted are the contours of the best fitting MGE models.

Using the different broad-band images we constructed color maps.
No color differences are seen between the different components (bulge, outer 
disk, nuclear disk). From a comparison with stellar population
models we conclude that the nuclear disk must be at least $\sim 5$ Gyr old,
and can not have a stellar population too different from that of the bulge.

\section Kinematics and Dynamical modeling

We have obtained long-slit spectroscopy along the major and minor axes
of NGC 4342 with the William Herschel Telescope (hereafter WHT) at La Palma. 
In addition, we obtained HST {\tt FOS} spectra with the circular $0.3''$ 
aperture at 7 positions in the central region, mainly 
along the nuclear disk. The rotation velocities and velocity dispersions along
the major axis are shown in Fig. 2.

\figureps[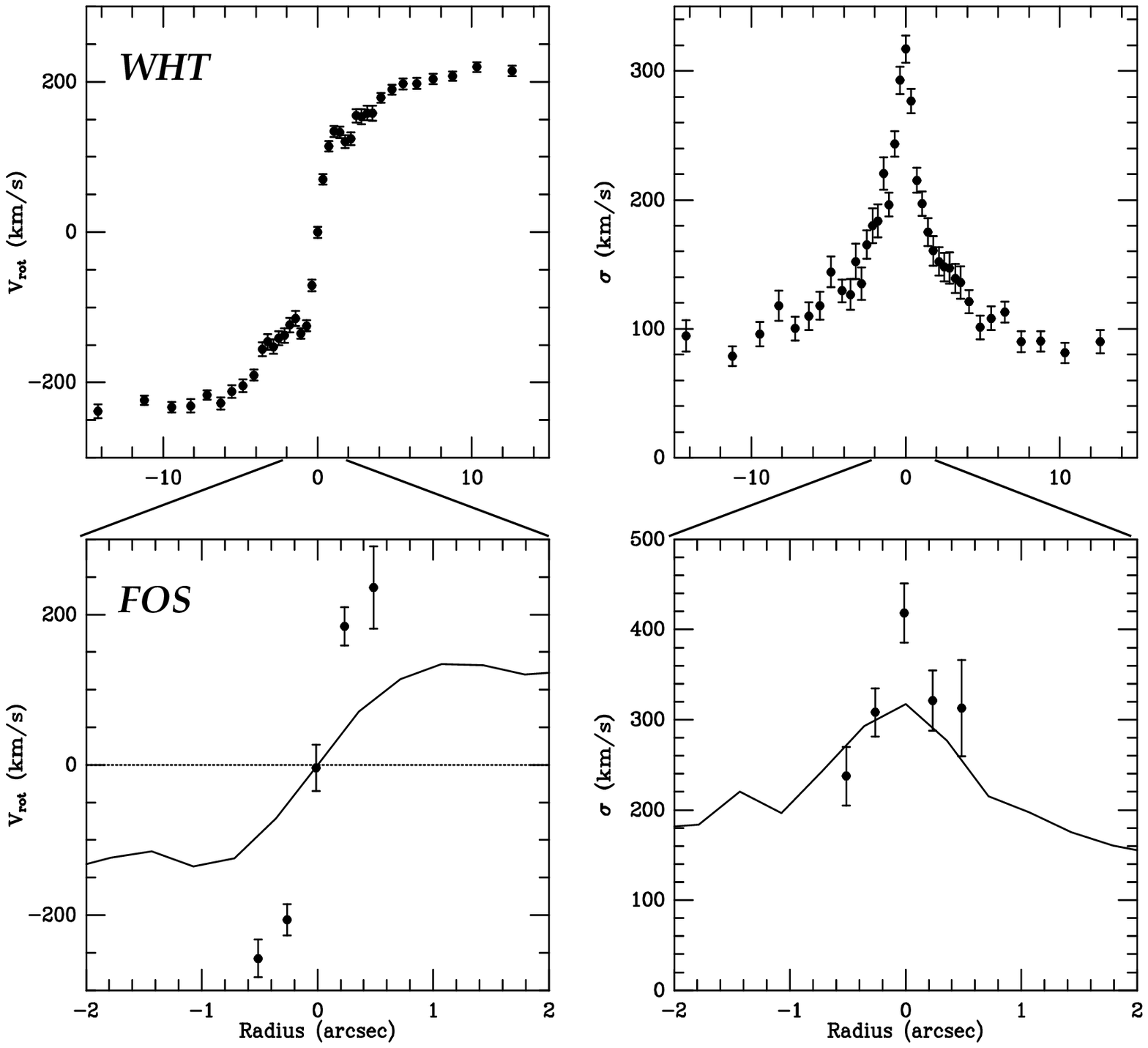,.75\hsize] 2. Rotation velocities and velocity 
dispersions along the major axis of NGC 4342 as obtained with the WHT and 
FOS (solid dots). The solid lines in the lower panel are the WHT 
measurements plotted for comparison.

We find a strong central increase in velocity dispersion.
The ground based spectra (seeing FWHM $\sim 0.8''$) give $\sigma_0 = 320
\kms$, which increases to $\sigma_0 = 418 \kms$ when observed with the 
{\tt FOS}.
In addition, the {\tt FOS} spectra reveal an enormously steep central rotation 
curve, with $V_{\rm rot} \sim 200 \kms$ at $0.25''$ from the center 
(corresponding to $18\pc$ at a distance of $15 \Mpc$). Clearly, both the 
rotation velocity and the velocity dispersions indicate very high central 
densities.

To investigate whether NGC 4342 harbors a massive BH, we 
constructed dynamical models using the MGE model discussed above as
luminosity density. Under the assumptions of constant M/L, that we observe 
NGC 4342 edge-on, and that the distribution function (DF) $f = f(E,L_z)$, we 
have solved the Jeans Equations. After line-of-sight projection of the 
velocity moments, we mimic the observations 
by taking PSF convolution and pixel binning into account.

\figureps[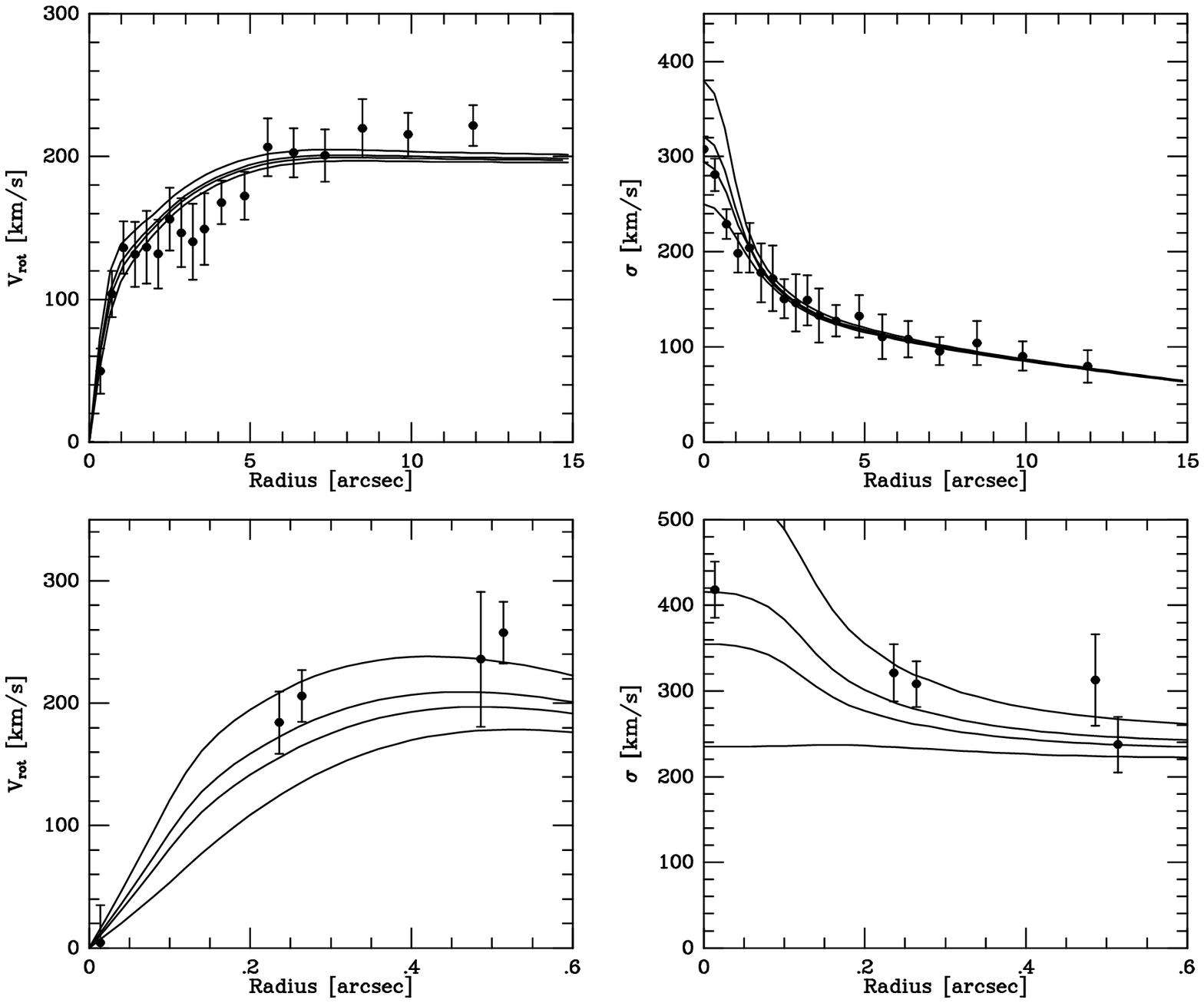,.8\hsize] 2. Rotation velocities and velocity
dispersions (folded) measured with the WHT and the FOS with overplotted
four different models that only differ in BH mass: 0, $3\times 10^8$,
$5\times 10^8$, and $1\times 10^9\Msun$. 

The results are shown in Fig. 3, where we have plotted the folded data 
together with four models 
that only differ in the mass of the central BH: $M_{\rm BH} = 0$, 
$3\times 10^8$, $5\times 10^8$, and $1\times 10^9\Msun$. Although there are 
several small discrepancies, models with $M_{\rm BH} = $3-6$\times 
10^8\Msun$ fit the observations moderately well.

\section Conclusions

The E/SO galaxy NGC 4342 harbors, in addition to its outer disk,
a bright, nuclear disk. Color images based on optical broad-band 
images obtained with {\tt WFPC2} reveal no color differences between any
of the subcomponents. Unless there is a conspiracy between metallicity and 
age, the age differences between any of the components is less than $\sim 3$
Gyr.

NGC 4342 is a small, faint elliptical (it is 3.5 magnitude fainter than
the morphologically similar NGC 3115).  We have shown here that it has
an extremely large central velocity dispersion ($>400 \kms$), as well
as an enormously steep central rotation curve. Axisymmetric Jeans modeling
suggests the presence of a massive BH of a few times $10^8 \Msun$.
However, these models are based on the assumption that the DF depends
only on the two classical integrals of motion, thereby neglecting the
possibility of strong radial anisotropy. It is well known that radial
orbits in the center of a galaxy can result in large central velocity
dispersions. This may therefore elude the requirement of a central BH in 
NGC 4342. However, the strong rotation in the center constrains
the amount of radial anisotropy, and it seems unlikely that a model 
without BH can fit both the central velocity dispersions and the steep rotation
curve. We are nevertheless in the process of constructing fully general, 
three integral dynamical models, using the technique described by  N. Cretton 
in these proceedings. Not only will this put more stringent constraints on the 
mass of the BH, but it will also allow an investigation of differences
in velocity anisotropy between the different components.

About a dozen galaxies are now known as candidates for harboring a massive 
BH. Kormendy \& Richstone (1995) suggested that there
appears to be a correlation between bulge mass and BH mass such
that $\langle \log (M_{\rm BH}/M_{\rm bulge})\rangle \sim -2.5 \pm 0.2$.
If, from the three integral modeling, the BH mass of NGC 4342 is confirmed 
to be a few times $10^8\Msun$, this is, after the Milky Way, the second
galaxy to strongly deviate
from this correlation ($\log (M_{\rm BH}/M_{\rm bulge}) = -1.4 \pm 0.3$).
Clearly, a larger sample is required to investigate this in more
detail. Galaxies with nuclear disks are particularly suited for this,
as is evident from the fact that three of the BH candidate galaxies 
(NGC 3115, NGC 4342, and NGC 4594) harbor such structures.

\references 

Kormendy J., Richstone D., 1995, ARA\&A, 33, 581

Lauer T.R., {\it et al.}, 1995, AJ, 110, 2622 

Monnet G., Bacon R., Emsellem E., 1992, A\&A, 253, 366

van den Bosch F.C., \etal, 1994, AJ, 108, 1579

van den Bosch F.C., de Zeeuw P.T., 1996, MNRAS, 283, 381

\bye